\documentclass[reprint, superscriptaddress, twocolumn, amsmath, amssymb, prx]{revtex4-2}

\usepackage{graphicx}
\usepackage{dcolumn} 
\usepackage{bm}
\usepackage[colorlinks=true, citecolor=magenta, linkcolor=magenta]{hyperref}

\usepackage{mathrsfs}
\usepackage{color}
\usepackage{soul}
\usepackage{braket}

\usepackage{lipsum}
\usepackage{xargs}
\usepackage{setspace}

\usepackage{float}
\usepackage{booktabs}

\usepackage{soul}
\usepackage{color}

\definecolor{orange}{rgb}{1.0, 0.5, 0.0}
\definecolor{violet}{rgb}{0.78,0.08, 0.52}
\definecolor{green}{rgb}{0.11, 0.35, 0.02}
\definecolor{bluebell}{rgb}{0.64, 0.64, 0.82}
\definecolor{capri}{rgb}{0.0, 0.45, 0.73}

\begin{document}

\title{Quantum enhancement of spoofing detection with squeezed states of light}

\def\ARL{DEVCOM Army Research Laboratory, Adelphi, MD 20783}
\def\UCH{Departamento de F\'isica, Facultad de Ciencias, Universidad de Chile, Santiago, Chile}

\author{Tom\'as P. Espinoza}
  \email{tp.espinozam@gmail.com}
  \affiliation{\UCH}

\author{Sebastian C. Carrasco}
  \email{seba.carrasco.m@gmail.com}
  \affiliation{\ARL}

\author{Jos\'e Rogan}
    \affiliation{\UCH}
    
\author{Juan Alejandro Valdivia}
    \affiliation{\UCH}

\author{Vladimir S. Malinovsky}
  \affiliation{\ARL}

\date{\today}

\begin{abstract}
We employ quantum state discrimination theory to establish the ultimate limit for spoofing detection in electromagnetic signals encoded with random quantum states. Our analysis yields an analytical expression for the optimal bound, which we prove can be achieved using a pair of coherent states. Notably, the quantum enhancement persists regardless of photon number, thereby removing the previous constraint to single-photon sources. This breakthrough unlocks new experimental possibilities using standard laser sources. Furthermore, we explore the encoding of squeezed states and demonstrate that the detection probability approaches 100\% when the spoofer's capability is restricted to coherent state generation.
\end{abstract}

\maketitle

\section{Introduction}

Quantum state discrimination theory~\cite{helstrom1969quantum, YuenIEEETIT1975} is a fundamental part of quantum mechanics and its technological applications~\cite{berman2011principles, BaeJPAMT2015}, such as quantum computing~\cite{SatyajitQIP2018}, quantum sensing~\cite{ZhuangPRL2017}, and quantum information science~\cite{BergouPRL2013, NakahiraPRA2015, SantraNJP2019, FieldsPRA2022}. It establishes that it is impossible to discriminate a set of non-orthogonal quantum states from each other with absolute precision~\cite{helstrom1969quantum, YuenIEEETIT1975}. At the most, one could successfully differentiate between states with a probability that depends on their overlap (in the case of pure quantum states). This has radical consequences. If quantum state discrimination were always perfect, it would imply impossibilities such as instantaneous communication via quantum entanglement~\cite{GisinPLA1998, BaePRL2011} (contradicting the no-signaling principle) or cloning quantum states~\cite{WoottersN1982, CheflesJPAMG1998} (contradicting the no-cloning theorem). Thus, the impossibility of perfect quantum state discrimination is closely related to other fundamental results of quantum mechanics, and one can regard it as a fundamental, yet implicit, postulate of quantum theory~\cite{BaeJPAMT2015}. 

These intriguing features have motivated applications of quantum mechanics in security and have given rise to new fields such as quantum cryptography~\cite{GisinRMP2002, PortmannRMP2022}. In the past decade, it has been pointed out that those quantum effects can enhance spoofing detection with respect to the limitations that classical physics imposes~\cite{MalikAPL2012, WilliamsPRA2016, ZhaoOE2022, OhPRL2023}. More recently, Blakely et al.~\cite{BlakelyPRR2022} showed that the most simple model of quantum spoofing exhibits enhancement in the probability of detecting spoofing. That enhancement comes from the impossibility of perfectly discriminating two non-orthogonal states, and its main application would be to detect the spoofing of a radar signal~\cite{Schleher, BlakelyPRR2022, BlakelyPRR2024}. In radar, one party tries to gain information about the other's position, speed, and orientation, to name a few, by analyzing the reflection of an electromagnetic signal. While the other, the spoofer, could attempt to provide false information by creating a fictitious reflection. One limitation in the model of Ref. ~\cite{BlakelyPRR2022} is that the quantum advantage exists only within the limits of a low number of photons. Thus limiting its applicability due to the difficulty of generating low photon pulses and distinguishing them from the background noise~\cite{SenellartNN2017}.

The model consists of a transmitter-receiver pair that chooses one of two non-orthogonal quantum states. Then, the transmitter encodes the state in an electromagnetic signal. A third party, the spoofer, could intercept the signal and send a new one to the receiver. The spoofer has a delicate task if it wants to be undetected. It needs to know which state the transmitter-receiver-pair chose. Thus, the spoofer must discriminate between the two possible quantum states, which the spoofer cannot do perfectly. The drawback is that the receiver must also measure the quantum state to check if it is the one agreed upon, which cannot be done faultlessly. In this model, that interplay is what generates but also limits the quantum enhancement. 

We extend the protocol of Ref.~\cite{BlakelyPRR2022} to explore the fundamental quantum limits of spoofing detection, considering arbitrary quantum states as transmitter-receiver pairs. This generalization surpasses the binary-phase shift keying scenario of~\cite{BlakelyPRR2022}, enabling a more comprehensive analysis. Our key findings reveal that the success probability of spoofing detection solely depends on the overlap between the quantum states, allowing us to derive an upper bound for this probability. Furthermore, we identify the optimal coherent states for the transmitter-receiver, which achieve this bound at a critical photon number and demonstrate quantum enhancement over classical success probabilities. Notably, our approach eliminates the need for single-photon sources in proof-of-principle experiments; thus, it opens the door for a new regime for experimental demonstrations, paving the way for experimental advancements and further development of spoofing detection technology.

At first glance, our claim that advantages persist at higher photon numbers may appear counterintuitive, as experiments typically demonstrate that the probability of error in quantum state discrimination decreases with increasing photons~\cite{CookN2007, WittmannPRL2008, FerdinandNPJQI2017}. However, these experiments usually involve discriminating between states with decreasing overlap as the photon number increases. In contrast, our approach optimizes quantum spoofing detection by identifying states with a specific optimal overlap value independent of photon number. As a result, the probability of successful discrimination remains constant, unaffected by the number of photons.

Our findings reveal that multiple pairs of quantum states, including nonclassical states, can saturate the bound. Although nonclassical states offer no advantage over coherent states in this context, their generation is significantly more challenging~\cite{SchnabelPR2017}. Therefore, we investigate a scenario where the transmitter-receiver pair has access to nonclassical states, while the spoofer is limited to coherent states. In this restricted scenario, we demonstrate that the success probability bound can be surpassed. As a specific example, we consider the transmitter-receiver pair utilizing squeezed states of light~\cite{WallsN1983, SlusherPRL1985, SchnabelPR2017}. Our analysis shows that the probability of detecting spoofing approaches unity, limited only by the experimental capabilities of generating nonclassical states, such as squeezed states.

\section{Quantum limits for spoofing two arbitrary quantum states}

\begin{figure}[H]
  \centering
  \includegraphics[scale=1.0]{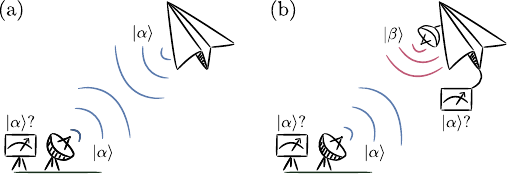}
  \caption{\label{fig:cartoon} Possible scenarios and quantum states. (a) shows the scenario where a quantum state $\ket{\alpha}$ is encoded in a signal and transmitted. Then, an object (represented by an airplane) reflects the signal to the receiver. (b) shows the alternative scenario where a spoofer intercepts the transmitted signal and creates a false reflection. The quantum state encoded in the spoof signal depends on whether the spoofer concluded that the state was $\ket{\alpha}$ or $\ket{\beta}$. In the image, the spoofer concluded that it was $\ket{\beta}$ and encoded it in the spoof signal. The receiver attempts to distinguish between both scenarios based on the received signal.
  }
\end{figure}

Let's consider the scenario illustrated in Fig.~\ref{fig:cartoon}(a). A transmitter randomly chooses a quantum state $\ket{\alpha}$ or $\ket{\beta}$ with equal probability. Next, the transmitter emits the coherent state encoded in an electromagnetic signal~\cite{ou2017quantum} (see Fig. \ref{fig:cartoon}(a)). Then, the signal reflects from a target and goes to the receiver, which also knows the chosen quantum state. Upon arrival, the receiver can get information regarding the target position and speed by analyzing the arrival of the reflecting signal and the Doppler shift, among other properties. Nevertheless, there is a chance that the target chooses to spoof the signal by emitting a false signal, providing misleading information (see Fig. \ref{fig:cartoon}(b)).

The receiver can detect if the signal is a truly reflected pulse by checking if the quantum state of the received signal is the same as that transmitted originally. So, if the spoofer wants to go unnoticed, it has to create a pulse that is an exact copy of the original (as in Fig. \ref{fig:cartoon}(b)). To create that copy, it must discriminate between the states $\ket{\alpha}$ or $\ket{\beta}$ that can be emitted by the transmitter. The measurement is able to successfully discriminate the states with probability
\begin{equation} \label{eq:gamma}
    \gamma = \frac{1}{2} + \frac{1}{2} \sqrt{1 - |\tau|^2} \, ,
\end{equation}
where $\tau = \braket{\alpha| \beta}$. This is called the Helstrom bound~\cite{helstrom1969quantum}.

When the reflected (or spoofed) pulse arrives at the receiver, the receiver must determine whether there has been spoofing. To do that, it has to choose between two hypotheses. The first one, $H_1$, is that the pulse is a true reflection of the original pulse, which we assume to be $\ket{\alpha}$. The density operator that describes this hypothesis is given by
\begin{equation}
    \hat{\rho}_1 = \ket{\alpha}\!\bra{\alpha} \, .
\end{equation}
The second hypothesis, $H_2$, holds that the target side spoofed the pulse. The spoofer optimally discriminates the quantum state with probability $\gamma$. Therefore, the density operator representing this hypothesis is mixed and is described by
\begin{equation}
    \hat{\rho}_2 = \gamma \ket{\alpha}\!\bra{\alpha} + ( 1 - \gamma ) \ket{\beta}\!\bra{\beta} \, .
\end{equation}
If the prior probability of $H_2$ is $p$, the maximum probability of successfully discriminating a true reflection from a spoof is
\begin{equation} \label{eq:general_success_prob}
    P_s = \dfrac{1}{2} +  \dfrac{1}{2} || p \hat{\rho}_2 - ( 1 - p ) \hat{\rho}_1 || \, ,
\end{equation}
where $||\cdot||$ is the trace norm, which is defined as the sum of the absolute values of the eigenvalues. 

The eigenvalues $\eta$ of $p \hat{\rho}_2 - ( 1 - p ) \hat{\rho}_1$ satisfy
\begin{equation}
    \begin{vmatrix}
        p \gamma - (1 - p) - \eta & \left[p \gamma - (1 - p)\right] \braket{\alpha | \beta} \\
        p (1-\gamma) \braket{\beta|\alpha} & p (1-\gamma) - \eta
    \end{vmatrix} = 0 \, ,
\end{equation}
and are given by
\begin{equation} \label{eq:eigenvals}
    \eta_\pm = \frac{1}{2} - p \pm \chi \, ,
\end{equation}
where $\chi = \sqrt{\left(1/2 - p \right)^2 - p ( 1 - \gamma)(p\gamma - 1 + p)  (2\gamma - 1)^2}$. The probability of success in detecting the spoofer is then
\begin{equation}
    P_s = \dfrac{1}{2} ( 1 + |\eta_+| + |\eta_-|) \, .
\end{equation}
As $\eta_+$ is always positive, the probability of success depends on the sign of $\eta_-$. Indeed,
\begin{equation} \label{eq:universal}
    P_s(\gamma) =
    \begin{cases}
			\dfrac{1}{2} + \chi & \text{if $p < (1 + \gamma)^{-1}$} \, ,\\
            p & \text{otherwise} \, .
    \end{cases} 
\end{equation}
If $p > (1 + \gamma)^{-1}$, then $\eta_-$ is positive, and the probability of success is $P_s = p$, which matches the classical limit (where the states can be unambiguously discriminated, and $\gamma=1$). If $p < (1 + \gamma)^{-1}$, $\eta_-$ is negative and the probability of success $P_s$ differs from the classical limit.

\begin{figure}
  \centering
  \includegraphics[]{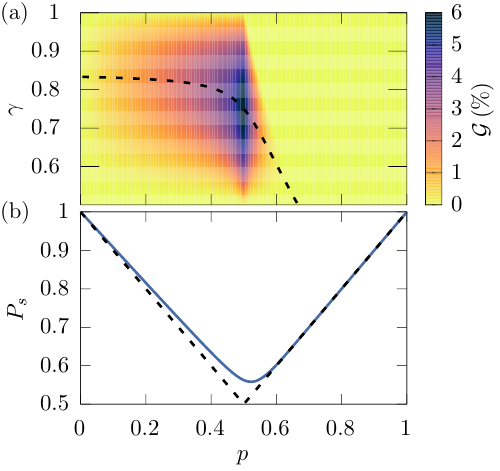}
    \caption{\label{fig:fig2} Quantum spoofing results for a pair of arbitrary quantum states. (a) Quantum enhancement of the success probability in detecting the spoofer, $\mathcal{G}$, as a function of the prior probability of spoofing $p$ and the probability of discriminating the two states $\gamma$. The dashed line shows the optimal value of $\gamma$. (b) Optimal bound for the success probability as a function of the prior probability $p$ for two arbitrary quantum states (solid line) and the classical limit (dashed line) where perfect state discrimination is possible, thus $\gamma=1$.
  }
\end{figure}

To find the maximum value of the success probability, we need to optimize $\chi$ in Eq. \eqref{eq:universal}, which is a function of $\gamma$. Taking $\partial \chi^2/\partial \gamma = 0$ (which shares the maximum we are looking for with $\chi$), we obtain that the optimal value of $\gamma$ should satisfy the following equation
\begin{equation}
    16 p \gamma^3 - 12 (p+1) \gamma^2 + (16 - 6 p) \gamma + 4p - 5 =0 \, .
\end{equation}
Solving this equation, we find the optimal discrimination probability between two arbitrary states $\ket{\alpha}$ and $\ket{\beta}$ to be 
\begin{equation} \label{eq:gamma_opt}
    \gamma_\text{opt} = \dfrac{p+3 - \sqrt{33p^2-34p+9}}{8p} \, ,
\end{equation}
and therefore, according to Eq. \eqref{eq:gamma}, the optimal overlap is 
\begin{equation} \label{eq:overlap_squared}
    |\tau_\text{opt}|^2 = \frac{26 p - 13 p^2-9 + 3(1-p) \sqrt{33p^2-34p+9} }{8 p^2} \, .
\end{equation}
The equations \eqref{eq:universal}, \eqref{eq:gamma_opt} and \eqref{eq:overlap_squared} are the main results of this work, which allow us to find all pairs of quantum states maximizing the success probability of spoofing detection. Together, these equations define a universal bound for optimal quantum spoofing detection.  

Fig.~\ref{fig:fig2}(a) shows the quantum enhancement of the probability of detecting spoofing, given by the success probability increase with respect to the classical limit, $\mathcal{G} = P_s(\gamma) - P_s(\gamma=1)$, as a function of $p$ and $\gamma$. In a dashed line, we plot the optimal discrimination probability, $\gamma_\text{opt}$. We observe that our solution maximizes the quantum enhancement and that the maximum is surprisingly wide, demonstrating the solutions' robustness. In Fig. \ref{fig:fig2}(b), we plot the optimal success probability of spoofing detection (solid line) as a function of the prior probability, $p$, and compare it with the classical limit (dashed line). Using the expression for the optimal value of the successful discrimination probability, $\gamma_\text{opt}$, in Eq.~\eqref{eq:gamma_opt} and the success probability in Eq.~\eqref{eq:universal} we find that a quantum enhancement exist only if the prior probability $p < 2/3$. As we can see, the quantum advantage is maximum around $p \approx 1/2$.

\section{quantum bound of spoofing detection: coherent states case}

To attain the bound, we modify the strategy of \cite{BlakelyPRR2022}, where the states $\ket{\alpha}$ and $\ket{-\alpha}$ (binary-phase shift keying) were used by the transmitter-receiver. Here we propose to use the coherent states $\ket{\alpha} = \ket{\sqrt{n} \exp(i \phi/2)}$ and $\ket{\beta} = \ket{\alpha^*}$, where $n$ is the number of photons. The idea is that the phase $\phi$ will allow us to control the overlap 
\begin{equation} \label{eq:cs_overlap}
    |\tau|^2=|\braket{\alpha|\beta}|^2=\exp(-4n \sin^2\phi/2) \, ,    
\end{equation}
and consequently, the state discrimination probability, $\gamma$, and the success probability of the spoofing detection, $P_s$.

Using Eq. \eqref{eq:cs_overlap}, we solve for the optimal phase given by
\begin{equation} \label{eq:phi_opt}
    \phi_\text{opt} = 2\arcsin\left( \sqrt{- \dfrac{\log |\tau_\text{opt}|}{2n}} \right) \, .
\end{equation}
In the limit of a large number of photons, $\phi_\text{opt}$ scales proportionally to $1/\sqrt{n}$ due to the fact that the distribution width of a coherent state scales as $\sqrt{n}$. For small values of $n$, $\phi_\text{opt}$ becomes undefined according to Eq. \eqref{eq:phi_opt} since the probability distributions of $\ket{\alpha}$ and $\ket{\beta}$ are too close to the origin in phase space and, thus, too close to each other. In this case, the optimal phase is $\phi_\text{opt}=\pi$, as it minimizes the overlap. For $p=1/2$, the critical number of photons upon $\phi_\text{opt}$ ceases to be $\pi$ is $\log(4/3)/4 \approx 0.072$ (the minimum number of photons where Eq. \eqref{eq:phi_opt} is well defined), as the optimal overlap in that case is $|\tau_\text{opt}|^2=3/4$.

\begin{figure}
  \centering
  \includegraphics[]{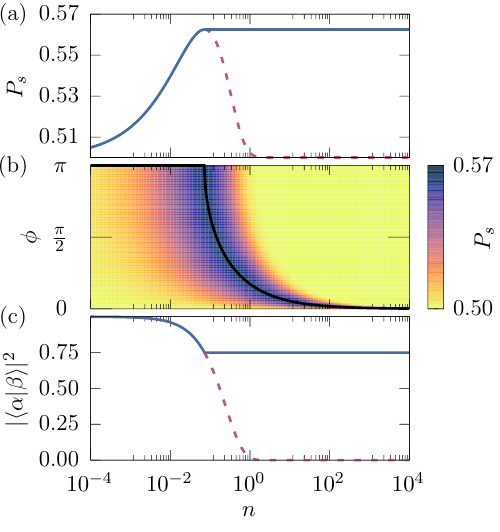}
  \caption{\label{fig:fig3} Quantum spoofing results for a pair of coherent quantum states if the prior probability is $p=0.5$ (spoofing is as likely as not spoofing). (a) Success probability of detecting spoofing as a function of the number of photons $n$. The dashed line corresponds to the case $\phi=\pi$, and the solid line demonstrates the proposed scenario when we adjust the state overlap by optimizing $\phi$. (b) Probability of detecting spoofing as a function of the phase, $\phi$, and the number of photons, $n$. The solid line shows the optimal phase, $\phi_\text{opt}$. (c) The coherent state overlap squared vs the number of photons; the dashed line corresponds to $\phi=\pi$, and the solid line corresponds to the overlap generated by the optimal phase, $\phi_\text{opt}$.
  }
\end{figure}

We summarize our results for coherent states in Fig.~\ref{fig:fig3}. Fig.~\ref{fig:fig3}(a) shows that given a prior probability, there is a critical number of photons upon which the case $\phi=\pi$ ceases to be optimal. Beyond that number of photons, the success probability saturates the bound defined by Eq. \eqref{eq:universal} when the phase is adjusted according to \eqref{eq:phi_opt}, and it becomes independent on the averaged number of photons, $n$.

Fig.~\ref{fig:fig3}(b) shows the success probability as a function of the number of photons, $n$, and the phase, $\phi$. The solid line shows the optimal phase, which is $\pi$ for small values of $n$. For larger values of $n$, it starts decreasing as $1/\sqrt{n}$. As $n$ increases, this becomes necessary to keep the overlap constant as Eq.~\eqref{eq:cs_overlap} demands. 
Here, it arises for similar reasons, as the minimum phase scale allows state discrimination. Fig.~\ref{fig:fig3}(c) shows the comparison between the overlap obtained by optimizing the phase, $\phi$,  and the overlap obtained for $\phi=\pi$. The results are identical when the number of photons is below the critical value $n=0.072$. When the number of photons surpasses the critical value, the overlap converges to the optimal value $|\tau_\text{opt}|^2=3/4$, which can be obtained from Eq. \eqref{eq:overlap_squared}.

In summary, we show that it is possible to saturate the quantum bound for quantum spoofing detection using coherent states upon a certain number of photons.

\section{Detecting spoofing with squeezed states}

An alternative way to attain the maximum bound for the probability of success in detecting a spoofer is to utilize a pair of squeezed states. The only condition that needs to be fulfilled is given by Eq.~\eqref{eq:overlap_squared}, which shows that the only requirement is to ensure the states have the optimal overlap. A squeezed state $\ket{\Psi}$ is defined as~\cite{berman2011principles}
\begin{equation}
    \ket{\Psi(\alpha, \zeta)} = \hat{D}(\alpha) \hat{S}(\zeta) \ket{0} \, ,
\end{equation}
where $\hat{D}(\alpha)$ is the displacement operator
\begin{equation}
    \hat{D}(\alpha) = \exp( \alpha \hat{a}^\dagger - \alpha^* \hat{a} ) \, ,
\end{equation}
and $\hat{S}(\zeta)$ is the squeezing operator
\begin{equation}
    \hat{S}(\zeta) = \exp\Big( \dfrac{1}{2} ( \zeta^* \hat{a}^2 - \zeta \hat{a}^{\dagger2} ) \Big) \, .
\end{equation}
Here $\alpha = \sqrt{n} e^{i\phi/2}$ and $\zeta = r e^{i \theta}$ are complex numbers, which define the position in phase space of the state $(\sqrt{2n} \, \cos(\phi/2), \sqrt{2n} \, \sin(\phi/2))$, $r$ is the squeezing coefficient, and $\theta$ is the squeezing phase (related to the squeezing angle in phase space). The overlap between the two squeezed 
states can be adjusted by these parameters to be the optimal one. 

Although squeezed states do not provide any advantage by themselves, they are harder to produce in an optical setup. For instance, if the spoofer is inside a moving object such as an airplane, it would not be that easy to have the necessary optics inside. Thus, it makes sense to consider the case where the spoofer cannot produce squeezed states. That changes the hypothesis previously discussed in the following way: after successfully discriminating between the two squeezed states $ \ket{\varphi} = \ket{\Psi(\alpha, \zeta)}$ and $\ket{\xi} = \ket{\Psi(\alpha^*, \zeta^*)}$ with probability $\gamma$, the spoofer has to use the coherent states that maximize the overlap with the squeezed coherent state to respond. 

Suppose the original state encoded in the pulse is $\ket{\varphi}$, the hypothesis $H_1$ is then represented by the density operator
\begin{equation}
    \hat{\rho}_1 = \ket{\varphi} \! \bra{\varphi}
\end{equation}
and the hypothesis $H_2$ is described by
\begin{equation}
    \hat{\rho}_2 = \gamma \ket{\alpha} \! \bra{\alpha} + ( 1 - \gamma ) \ket{\alpha^*} \! \bra{\alpha^*} \, ,
\end{equation}
as the spoofer has probability $\gamma$ of concluding that the state encoded in the pulse was the squeezed state $\ket{\varphi} = \ket{\Psi(\alpha, \zeta)}$, and attempting to spoof with the state $\ket{\alpha} = \ket{\Psi(\alpha, 0)}$, which maximizes the overlap with the squeezed states. Similarly, the spoofer has probability $1-\gamma$ to conclude that the state was $\ket{\xi} = \ket{\Psi(\alpha^*, \zeta^*)}$ and respond with $\ket{\alpha^*} = \ket{\Psi(\alpha^*, 0)}$.

\begin{figure}[tb]
  \centering
  \includegraphics[width=1\columnwidth]{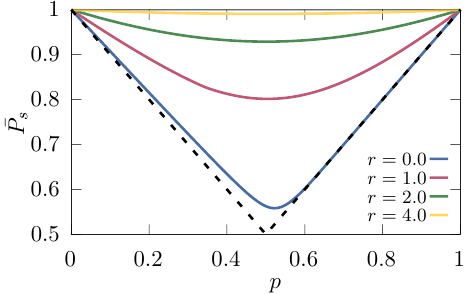}
  \caption{\label{fig:optimal_success_prob2} Maximum success probability as function of the prior probability, $p$, for $n=100$ photons. The dashed line corresponds to the classical limit, and each solid line corresponds to different values of squeezing $r$.
  }
\end{figure}

These hypotheses and Eq.~\eqref{eq:general_success_prob} give rise to a success probability $\bar P_s$ without the upper bound introduced in the previous section. The success probability can also be derived analytically, as shown in the Appendix. In Fig.~\ref{fig:optimal_success_prob2}, we corroborate this by plotting the success probability $\bar P_s$ as a function of the prior probability, $p$, for fixed values of the squeezing coefficient $r$ and an average number of photons $n=100$. By changing $\phi$ and $\theta$, we vary the position of the states in phase space and the direction of the squeezing in each calculation to maximize the success probability. We observe that the upper bound for $\bar P_s$ calculated in the previous section is broken when employing squeezed states. As we show, the higher the squeezing parameter, the higher the probability of detecting spoofing. Indeed, Fig.~\ref{fig:optimal_success_prob2} demonstrates that there is no upper bound for the success probability $\bar P_s$, and it is possible to detect spoofing with absolute accuracy for a sufficiently high squeezing parameter.
\begin{figure}[tb]
    \centering    
    \includegraphics[width=\columnwidth] {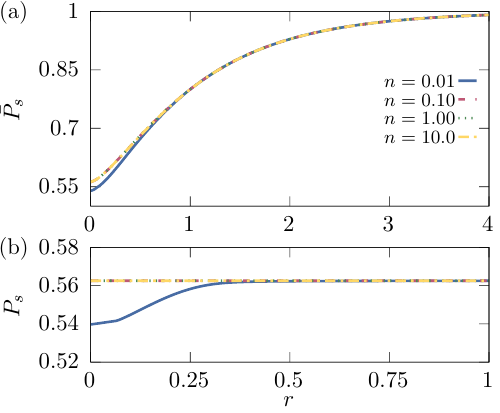}
    \caption{Success probability as a function of the squeezing coefficient for a various number of photons while the prior probability is $p=0.5$. (a) shows the success probability when the spoofer can only generate coherent states. (b) shows the results for the case where we remove that limitation on the spoofer capability. The states are optimized to simultaneously maximize the probability of success in both cases.}
    \label{fig:f5}
\end{figure}

It is constructive to consider a scenario when some spoofers can generate squeezing while others cannot. It turns out that it is possible to optimize the success probability for both cases simultaneously, independently on the ability of spoofers to generate squeezing. To demonstrate this, we plot in Fig.~\ref{fig:f5} the maximum success probability for both versions of the problem as a function of $r$ for different numbers of photons. To simultaneously optimize both cases, we vary $\phi$ and $\theta$ for each value $n$ and $r$. As Fig.~\ref{fig:f5}(a) shows, $\bar{P}_s$ converges to the unity as the squeezing increases independently of the number of photons. Indeed, if the number of photons is high enough, the curves coincide. We observe differences between the curves for a small number of photons as they coincide with the results from the previous section where the coherent states, $r=0$, were addressed. On the other hand, Fig.~\ref{fig:f5}(b) shows that this can be done while still saturating the success probability bound for the scenario where the spoofer can generate squeezing (and thus, the overlap must be the optimal one). In the limit $r=0$, the results coincide as well with those for coherent states discussed in the previous section. In particular, the success probability $P_s$ for $n=0.01$ photons does not saturate to the optimal value since the the number of photons is below the critical value of $0.072$. If the spoofer cannot generate squeezing, a squeezing coefficient of $r \approx 3.9$ is necessary to obtain $\bar{P}_s = 0.99$. That squeezing coefficient corresponds to $34$ dB of squeezing, which is outside of the current experimental capabilities~\cite{VahlbruchPRL2016}. We estimate that current experimental capabilities allow for $\bar{P}_s \approx 0.9$ employing 15 dB of squeezing, which is a substantial improvement with respect to the classical case. We attribute this result to the fact that for a high enough squeezing parameter, the overlap between the squeezed states and the coherent states approaches zero, making it possible to discriminate them with high probability; therefore, the receiver can perfectly identify a spoofer. 

\section{Conclusion and outlook}

We investigate a spoofing protocol that encodes arbitrary quantum states in an electromagnetic signal and derive the upper bound for spoofer detection probability. Our analysis reveals quantum enhancement in spoofing detection success probability, stemming from the impossibility of perfectly discriminating non-orthogonal quantum states. We demonstrate that a pair of coherent states can saturate this bound and determine the optimal states analytically. Furthermore, we show that this quantum enhancement persists for an arbitrarily large number of photons when using general coherent states for signal encoding. Notably, this result eliminates the strict requirement for single-photon sources, enabling a proof-of-principle experiment using standard laser sources to showcase quantum mechanics' role in improving spoofing detection~\cite{CookN2007, WittmannPRL2008, FerdinandNPJQI2017}.

Our work seeks to establish fundamental limits for quantum spoofing detection using the Helstrom bound framework, regarded as the most fundamental bound~\cite{CookN2007, WittmannPRL2008}. However, other sub-optimal bounds, such as direct photon counting~\cite{CookN2007} or unambiguous state discrimination~\cite{CheflesJPAMG1998}, may yield different limits for quantum spoofing detection.

Our results have significant implications for radar systems~\cite{Schleher, BlakelyPRR2022, BlakelyPRR2024}, particularly in scenarios where an airborne target emits spoof pulses to evade tracking by a ground-based radar. We investigate how the spoofer's ability to encode quantum states in the spoofing signal affects the probability of detection. For that purpose, we analyze the benefits of using squeezed states. Assuming only the transmitter-receiver can generate non-classical states like squeezed light~\cite{WallsN1983, SlusherPRL1985, SchnabelPR2017}, we show that the detection probability approaches unity, surpassing the coherent state case. With current experimental capabilities~\cite{VahlbruchPRL2016}, our estimate suggests achieving 90\% spoofing detection probability is feasible. However, if the spoofer can also generate squeezing, the advantage of squeezed states over coherent states diminishes as the photon number increases. Nevertheless, exploring more complex entangled states or modifying the encoding protocol could potentially harness quantum correlations to enhance spoofing detection further.

\begin{acknowledgments} 

This research was supported by DEVCOM Army Research Laboratory under Cooperative Agreement Number W911NF-24-2-0044 (SCC). JAV acknowledges support from AFOSR project FA9550-20-1-0189. JAV and JR Acknowledge partial support from ANID/Fondecyt grant N$^o$ 1240697 and N$^o$  1240655, respectively. The authors thank Jonathan N. Blakely for his helpful discussions of this work.

\end{acknowledgments} 

\medskip

\appendix

\renewcommand{\theequation}{A\arabic{equation}}
\renewcommand{\thefigure}{A\arabic{figure}}
\setcounter{figure}{0}
\setcounter{equation}{0}

\section{Eigenvalues calculation for the limited quantum resources scenario}

Analogously to Eq. \eqref{eq:general_success_prob}, the success probability $\bar P_s$ depends on the eigenvalues of the operator
%
\begin{multline}
    p \hat{\rho}_2 - ( 1 - p ) \hat{\rho}_1 = p \left( \gamma \ket{\alpha}\!\bra{\alpha} + ( 1 - \gamma ) \ket{\beta}\!\bra{\beta} \right) \nonumber\\
    - (1-p) \ket{\varphi}\!\bra{\varphi} \, . \nonumber
\end{multline}
%
Any eigenvector with a nonzero eigenvalue must be part of the subspace defined by $\ket{\alpha}$,  $\ket{\beta}$ and $\ket{\varphi}$. Thus, we can represent the previously introduced operator as a matrix in this subspace. The eigenvalues $\eta$ of that matrix satisfy
\begin{equation*}
    \begin{vmatrix}
        p \gamma - \eta & p \gamma \braket{\alpha|\beta} & p \gamma \braket{\alpha|\varphi} \\
        p (1 - \gamma) \braket{\beta|\alpha} & p (1 - \gamma) - \eta & p (1 - \gamma) \braket{\beta|\varphi} \\
        -(1 - p) \braket{\varphi|\alpha} & -(1 - p) \braket{\varphi|\beta} & -(1 - p) - \eta
    \end{vmatrix} = 0 \, ,
\end{equation*}
or equivalently,
\begin{equation*}
    a_3 \eta^3 + a_2 \eta^2 + a_1 \eta + a_0 = 0 \, ,
\end{equation*}
where
\begin{align}
    a_3 &= -1 \, , \nonumber\\
    a_2 &= 2p - 1 \, , \nonumber\\
    a_1 &= p ( -p\gamma + p\gamma^2 - p + 1) + p^2 \gamma (1-\gamma) |\braket{\alpha|\beta}|^2 \nonumber \\
    &- p(1-\gamma)(1-p) |\braket{\beta|\varphi}|^2 - p\gamma(1-p) |\braket{\alpha|\varphi}|^2 \, , \nonumber\\
    a_0 &= p^2 \gamma (1-\gamma) (1-p) \Big( |\braket{\alpha|\beta}|^2 + |\braket{\beta|\varphi}|^2 + |\braket{\alpha|\varphi}|^2 \nonumber \\
    &- 1 - 2 \, \mathfrak{Re} \{ \braket{\beta|\varphi} \braket{\varphi|\alpha} \braket{\alpha|\beta} \} \Big) \, . \nonumber
\end{align}
Finally, the overlap between squeezed coherent states can be evaluated using~\cite{mo1996displaced}
%

\begin{multline*}
\braket{\Psi(\alpha_1, \zeta_1)|\Psi(\alpha_2, \zeta_2)} \\ =
\dfrac{1}{
    \sqrt{\sigma_{21}}
    } \exp\left(\dfrac{\kappa_{21} \kappa_{12}^*}{2 \sigma_{21}} + \dfrac{ \alpha_2 \alpha_1^* - \alpha_2^* \alpha_1}{2}\right) \, ,
\end{multline*}
where
\begin{equation*}
    \sigma_{kl} = \cosh r_k \cosh r_l - e^{i(\theta_k - \theta_l)} \sinh r_k \sinh r_l \, ,
\end{equation*}
and
\begin{equation*}
    \kappa_{kl} = (\alpha_k - \alpha_l) \cosh r_k + (\alpha_k^* - \alpha_l^*) e^{i\theta_k} \sinh r_k \, .
\end{equation*}
%

\bibliography{refs}

\begin{thebibliography}{33}%
\makeatletter
\providecommand \@ifxundefined [1]{%
 \@ifx{#1\undefined}
}%
\providecommand \@ifnum [1]{%
 \ifnum #1\expandafter \@firstoftwo
 \else \expandafter \@secondoftwo
 \fi
}%
\providecommand \@ifx [1]{%
 \ifx #1\expandafter \@firstoftwo
 \else \expandafter \@secondoftwo
 \fi
}%
\providecommand \natexlab [1]{#1}%
\providecommand \enquote  [1]{``#1''}%
\providecommand \bibnamefont  [1]{#1}%
\providecommand \bibfnamefont [1]{#1}%
\providecommand \citenamefont [1]{#1}%
\providecommand \href@noop [0]{\@secondoftwo}%
\providecommand \href [0]{\begingroup \@sanitize@url \@href}%
\providecommand \@href[1]{\@@startlink{#1}\@@href}%
\providecommand \@@href[1]{\endgroup#1\@@endlink}%
\providecommand \@sanitize@url [0]{\catcode `\\12\catcode `\$12\catcode `\&12\catcode `\#12\catcode `\^12\catcode `\_12\catcode `\%12\relax}%
\providecommand \@@startlink[1]{}%
\providecommand \@@endlink[0]{}%
\providecommand \url  [0]{\begingroup\@sanitize@url \@url }%
\providecommand \@url [1]{\endgroup\@href {#1}{\urlprefix }}%
\providecommand \urlprefix  [0]{URL }%
\providecommand \Eprint [0]{\href }%
\providecommand \doibase [0]{https://doi.org/}%
\providecommand \selectlanguage [0]{\@gobble}%
\providecommand \bibinfo  [0]{\@secondoftwo}%
\providecommand \bibfield  [0]{\@secondoftwo}%
\providecommand \translation [1]{[#1]}%
\providecommand \BibitemOpen [0]{}%
\providecommand \bibitemStop [0]{}%
\providecommand \bibitemNoStop [0]{.\EOS\space}%
\providecommand \EOS [0]{\spacefactor3000\relax}%
\providecommand \BibitemShut  [1]{\csname bibitem#1\endcsname}%
\let\auto@bib@innerbib\@empty
\bibitem [{\citenamefont {Helstrom}(1969)}]{helstrom1969quantum}%
  \BibitemOpen
  \bibfield  {author} {\bibinfo {author} {\bibfnamefont {C.~W.}\ \bibnamefont {Helstrom}},\ }\bibfield  {title} {\bibinfo {title} {Quantum detection and estimation theory},\ }\href {https://doi.org/https://doi.org/10.1007/BF01007479} {\bibfield  {journal} {\bibinfo  {journal} {Journal of Statistical Physics}\ }\textbf {\bibinfo {volume} {1}},\ \bibinfo {pages} {231} (\bibinfo {year} {1969})}\BibitemShut {NoStop}%
\bibitem [{\citenamefont {Yuen}\ \emph {et~al.}(1975)\citenamefont {Yuen}, \citenamefont {Kennedy},\ and\ \citenamefont {Lax}}]{YuenIEEETIT1975}%
  \BibitemOpen
  \bibfield  {author} {\bibinfo {author} {\bibfnamefont {H.}~\bibnamefont {Yuen}}, \bibinfo {author} {\bibfnamefont {R.}~\bibnamefont {Kennedy}},\ and\ \bibinfo {author} {\bibfnamefont {M.}~\bibnamefont {Lax}},\ }\bibfield  {title} {\bibinfo {title} {Optimum testing of multiple hypotheses in quantum detection theory},\ }\href {https://doi.org/10.1109/tit.1975.1055351} {\bibfield  {journal} {\bibinfo  {journal} {IEEE Trans. Inform. Theory}\ }\textbf {\bibinfo {volume} {21}},\ \bibinfo {pages} {125} (\bibinfo {year} {1975})}\BibitemShut {NoStop}%
\bibitem [{\citenamefont {Berman}\ and\ \citenamefont {Malinovsky}(2011)}]{berman2011principles}%
  \BibitemOpen
  \bibfield  {author} {\bibinfo {author} {\bibfnamefont {P.~R.}\ \bibnamefont {Berman}}\ and\ \bibinfo {author} {\bibfnamefont {V.~S.}\ \bibnamefont {Malinovsky}},\ }\href@noop {} {\emph {\bibinfo {title} {Principles of laser spectroscopy and quantum optics}}}\ (\bibinfo  {publisher} {Princeton University Press},\ \bibinfo {year} {2011})\BibitemShut {NoStop}%
\bibitem [{\citenamefont {Bae}\ and\ \citenamefont {Kwek}(2015)}]{BaeJPAMT2015}%
  \BibitemOpen
  \bibfield  {author} {\bibinfo {author} {\bibfnamefont {J.}~\bibnamefont {Bae}}\ and\ \bibinfo {author} {\bibfnamefont {L.-C.}\ \bibnamefont {Kwek}},\ }\bibfield  {title} {\bibinfo {title} {Quantum state discrimination and its applications},\ }\href {https://doi.org/10.1088/1751-8113/48/8/083001} {\bibfield  {journal} {\bibinfo  {journal} {J. Phys. A: Math. Theor.}\ }\textbf {\bibinfo {volume} {48}},\ \bibinfo {pages} {083001} (\bibinfo {year} {2015})}\BibitemShut {NoStop}%
\bibitem [{\citenamefont {Satyajit}\ \emph {et~al.}(2018)\citenamefont {Satyajit}, \citenamefont {Srinivasan}, \citenamefont {Behera},\ and\ \citenamefont {Panigrahi}}]{SatyajitQIP2018}%
  \BibitemOpen
  \bibfield  {author} {\bibinfo {author} {\bibfnamefont {S.}~\bibnamefont {Satyajit}}, \bibinfo {author} {\bibfnamefont {K.}~\bibnamefont {Srinivasan}}, \bibinfo {author} {\bibfnamefont {B.~K.}\ \bibnamefont {Behera}},\ and\ \bibinfo {author} {\bibfnamefont {P.~K.}\ \bibnamefont {Panigrahi}},\ }\bibfield  {title} {\bibinfo {title} {Nondestructive discrimination of a new family of highly entangled states in {IBM} quantum computer},\ }\href {https://doi.org/10.1007/s11128-018-1976-9} {\bibfield  {journal} {\bibinfo  {journal} {Quantum Inf Process}\ }\textbf {\bibinfo {volume} {17}},\ \bibinfo {pages} {212} (\bibinfo {year} {2018})}\BibitemShut {NoStop}%
\bibitem [{\citenamefont {Zhuang}\ \emph {et~al.}(2017)\citenamefont {Zhuang}, \citenamefont {Zhang},\ and\ \citenamefont {Shapiro}}]{ZhuangPRL2017}%
  \BibitemOpen
  \bibfield  {author} {\bibinfo {author} {\bibfnamefont {Q.}~\bibnamefont {Zhuang}}, \bibinfo {author} {\bibfnamefont {Z.}~\bibnamefont {Zhang}},\ and\ \bibinfo {author} {\bibfnamefont {J.~H.}\ \bibnamefont {Shapiro}},\ }\bibfield  {title} {\bibinfo {title} {Optimum mixed-state discrimination for noisy entanglement-enhanced sensing},\ }\href {https://doi.org/10.1103/physrevlett.118.040801} {\bibfield  {journal} {\bibinfo  {journal} {Phys. Rev. Lett.}\ }\textbf {\bibinfo {volume} {118}},\ \bibinfo {pages} {040801} (\bibinfo {year} {2017})}\BibitemShut {NoStop}%
\bibitem [{\citenamefont {Bergou}\ \emph {et~al.}(2013)\citenamefont {Bergou}, \citenamefont {Feldman},\ and\ \citenamefont {Hillery}}]{BergouPRL2013}%
  \BibitemOpen
  \bibfield  {author} {\bibinfo {author} {\bibfnamefont {J.}~\bibnamefont {Bergou}}, \bibinfo {author} {\bibfnamefont {E.}~\bibnamefont {Feldman}},\ and\ \bibinfo {author} {\bibfnamefont {M.}~\bibnamefont {Hillery}},\ }\bibfield  {title} {\bibinfo {title} {Extracting information from a qubit by multiple observers: Toward a theory of sequential state discrimination},\ }\href {https://doi.org/10.1103/physrevlett.111.100501} {\bibfield  {journal} {\bibinfo  {journal} {Phys. Rev. Lett.}\ }\textbf {\bibinfo {volume} {111}},\ \bibinfo {pages} {100501} (\bibinfo {year} {2013})}\BibitemShut {NoStop}%
\bibitem [{\citenamefont {Nakahira}\ \emph {et~al.}(2015)\citenamefont {Nakahira}, \citenamefont {Kato},\ and\ \citenamefont {Usuda}}]{NakahiraPRA2015}%
  \BibitemOpen
  \bibfield  {author} {\bibinfo {author} {\bibfnamefont {K.}~\bibnamefont {Nakahira}}, \bibinfo {author} {\bibfnamefont {K.}~\bibnamefont {Kato}},\ and\ \bibinfo {author} {\bibfnamefont {T.~S.}\ \bibnamefont {Usuda}},\ }\bibfield  {title} {\bibinfo {title} {Generalized quantum state discrimination problems},\ }\href {https://doi.org/10.1103/physreva.91.052304} {\bibfield  {journal} {\bibinfo  {journal} {Phys. Rev. A}\ }\textbf {\bibinfo {volume} {91}},\ \bibinfo {pages} {052304} (\bibinfo {year} {2015})}\BibitemShut {NoStop}%
\bibitem [{\citenamefont {Santra}\ \emph {et~al.}(2019)\citenamefont {Santra}, \citenamefont {Muralidharan}, \citenamefont {Lichtman}, \citenamefont {Jiang}, \citenamefont {Monroe},\ and\ \citenamefont {Malinovsky}}]{SantraNJP2019}%
  \BibitemOpen
  \bibfield  {author} {\bibinfo {author} {\bibfnamefont {S.}~\bibnamefont {Santra}}, \bibinfo {author} {\bibfnamefont {S.}~\bibnamefont {Muralidharan}}, \bibinfo {author} {\bibfnamefont {M.}~\bibnamefont {Lichtman}}, \bibinfo {author} {\bibfnamefont {L.}~\bibnamefont {Jiang}}, \bibinfo {author} {\bibfnamefont {C.}~\bibnamefont {Monroe}},\ and\ \bibinfo {author} {\bibfnamefont {V.~S.}\ \bibnamefont {Malinovsky}},\ }\bibfield  {title} {\bibinfo {title} {Quantum repeaters based on two species trapped ions},\ }\href {https://doi.org/10.1088/1367-2630/ab2a45} {\bibfield  {journal} {\bibinfo  {journal} {New J. Phys.}\ }\textbf {\bibinfo {volume} {21}},\ \bibinfo {pages} {073002} (\bibinfo {year} {2019})}\BibitemShut {NoStop}%
\bibitem [{\citenamefont {Fields}\ \emph {et~al.}(2022)\citenamefont {Fields}, \citenamefont {Bergou}, \citenamefont {Hillery}, \citenamefont {Santra},\ and\ \citenamefont {Malinovsky}}]{FieldsPRA2022}%
  \BibitemOpen
  \bibfield  {author} {\bibinfo {author} {\bibfnamefont {D.}~\bibnamefont {Fields}}, \bibinfo {author} {\bibfnamefont {J.~A.}\ \bibnamefont {Bergou}}, \bibinfo {author} {\bibfnamefont {M.}~\bibnamefont {Hillery}}, \bibinfo {author} {\bibfnamefont {S.}~\bibnamefont {Santra}},\ and\ \bibinfo {author} {\bibfnamefont {V.~S.}\ \bibnamefont {Malinovsky}},\ }\bibfield  {title} {\bibinfo {title} {Optimal unambiguous discrimination of {Bell}-like states with linear optics},\ }\href {https://doi.org/10.1103/physreva.106.023706} {\bibfield  {journal} {\bibinfo  {journal} {Phys. Rev. A}\ }\textbf {\bibinfo {volume} {106}},\ \bibinfo {pages} {023706} (\bibinfo {year} {2022})}\BibitemShut {NoStop}%
\bibitem [{\citenamefont {Gisin}(1998)}]{GisinPLA1998}%
  \BibitemOpen
  \bibfield  {author} {\bibinfo {author} {\bibfnamefont {N.}~\bibnamefont {Gisin}},\ }\bibfield  {title} {\bibinfo {title} {Quantum cloning without signaling},\ }\href {https://doi.org/10.1016/s0375-9601(98)00170-4} {\bibfield  {journal} {\bibinfo  {journal} {Phys. Lett. A}\ }\textbf {\bibinfo {volume} {242}},\ \bibinfo {pages} {1} (\bibinfo {year} {1998})}\BibitemShut {NoStop}%
\bibitem [{\citenamefont {Bae}\ \emph {et~al.}(2011)\citenamefont {Bae}, \citenamefont {Hwang},\ and\ \citenamefont {Han}}]{BaePRL2011}%
  \BibitemOpen
  \bibfield  {author} {\bibinfo {author} {\bibfnamefont {J.}~\bibnamefont {Bae}}, \bibinfo {author} {\bibfnamefont {W.-Y.}\ \bibnamefont {Hwang}},\ and\ \bibinfo {author} {\bibfnamefont {Y.-D.}\ \bibnamefont {Han}},\ }\bibfield  {title} {\bibinfo {title} {No-signaling principle can determine optimal quantum state discrimination},\ }\href {https://doi.org/10.1103/physrevlett.107.170403} {\bibfield  {journal} {\bibinfo  {journal} {Phys. Rev. Lett.}\ }\textbf {\bibinfo {volume} {107}},\ \bibinfo {pages} {170403} (\bibinfo {year} {2011})}\BibitemShut {NoStop}%
\bibitem [{\citenamefont {Wootters}\ and\ \citenamefont {Zurek}(1982)}]{WoottersN1982}%
  \BibitemOpen
  \bibfield  {author} {\bibinfo {author} {\bibfnamefont {W.~K.}\ \bibnamefont {Wootters}}\ and\ \bibinfo {author} {\bibfnamefont {W.~H.}\ \bibnamefont {Zurek}},\ }\bibfield  {title} {\bibinfo {title} {A single quantum cannot be cloned},\ }\href {https://doi.org/10.1038/299802a0} {\bibfield  {journal} {\bibinfo  {journal} {Nature}\ }\textbf {\bibinfo {volume} {299}},\ \bibinfo {pages} {802} (\bibinfo {year} {1982})}\BibitemShut {NoStop}%
\bibitem [{\citenamefont {Chefles}\ and\ \citenamefont {Barnett}(1998)}]{CheflesJPAMG1998}%
  \BibitemOpen
  \bibfield  {author} {\bibinfo {author} {\bibfnamefont {A.}~\bibnamefont {Chefles}}\ and\ \bibinfo {author} {\bibfnamefont {S.~M.}\ \bibnamefont {Barnett}},\ }\bibfield  {title} {\bibinfo {title} {Quantum state separation, unambiguous discrimination and exact cloning},\ }\href {https://doi.org/10.1088/0305-4470/31/50/007} {\bibfield  {journal} {\bibinfo  {journal} {J. Phys. A: Math. Gen.}\ }\textbf {\bibinfo {volume} {31}},\ \bibinfo {pages} {10097} (\bibinfo {year} {1998})}\BibitemShut {NoStop}%
\bibitem [{\citenamefont {Gisin}\ \emph {et~al.}(2002)\citenamefont {Gisin}, \citenamefont {Ribordy}, \citenamefont {Tittel},\ and\ \citenamefont {Zbinden}}]{GisinRMP2002}%
  \BibitemOpen
  \bibfield  {author} {\bibinfo {author} {\bibfnamefont {N.}~\bibnamefont {Gisin}}, \bibinfo {author} {\bibfnamefont {G.}~\bibnamefont {Ribordy}}, \bibinfo {author} {\bibfnamefont {W.}~\bibnamefont {Tittel}},\ and\ \bibinfo {author} {\bibfnamefont {H.}~\bibnamefont {Zbinden}},\ }\bibfield  {title} {\bibinfo {title} {Quantum cryptography},\ }\href {https://doi.org/10.1103/revmodphys.74.145} {\bibfield  {journal} {\bibinfo  {journal} {Rev. Mod. Phys.}\ }\textbf {\bibinfo {volume} {74}},\ \bibinfo {pages} {145} (\bibinfo {year} {2002})}\BibitemShut {NoStop}%
\bibitem [{\citenamefont {Portmann}\ and\ \citenamefont {Renner}(2022)}]{PortmannRMP2022}%
  \BibitemOpen
  \bibfield  {author} {\bibinfo {author} {\bibfnamefont {C.}~\bibnamefont {Portmann}}\ and\ \bibinfo {author} {\bibfnamefont {R.}~\bibnamefont {Renner}},\ }\bibfield  {title} {\bibinfo {title} {Security in quantum cryptography},\ }\href {https://doi.org/10.1103/revmodphys.94.025008} {\bibfield  {journal} {\bibinfo  {journal} {Rev. Mod. Phys.}\ }\textbf {\bibinfo {volume} {94}},\ \bibinfo {pages} {025008} (\bibinfo {year} {2022})}\BibitemShut {NoStop}%
\bibitem [{\citenamefont {Malik}\ \emph {et~al.}(2012)\citenamefont {Malik}, \citenamefont {Magaña-Loaiza},\ and\ \citenamefont {Boyd}}]{MalikAPL2012}%
  \BibitemOpen
  \bibfield  {author} {\bibinfo {author} {\bibfnamefont {M.}~\bibnamefont {Malik}}, \bibinfo {author} {\bibfnamefont {O.~S.}\ \bibnamefont {Magaña-Loaiza}},\ and\ \bibinfo {author} {\bibfnamefont {R.~W.}\ \bibnamefont {Boyd}},\ }\bibfield  {title} {\bibinfo {title} {Quantum-secured imaging},\ }\bibfield  {journal} {\bibinfo  {journal} {Appl. Phys. Lett.}\ }\textbf {\bibinfo {volume} {101}},\ \href {https://doi.org/https://doi.org/10.1063/1.4770298} {https://doi.org/10.1063/1.4770298} (\bibinfo {year} {2012})\BibitemShut {NoStop}%
\bibitem [{\citenamefont {Williams}\ \emph {et~al.}(2016)\citenamefont {Williams}, \citenamefont {Britt},\ and\ \citenamefont {Humble}}]{WilliamsPRA2016}%
  \BibitemOpen
  \bibfield  {author} {\bibinfo {author} {\bibfnamefont {B.~P.}\ \bibnamefont {Williams}}, \bibinfo {author} {\bibfnamefont {K.~A.}\ \bibnamefont {Britt}},\ and\ \bibinfo {author} {\bibfnamefont {T.~S.}\ \bibnamefont {Humble}},\ }\bibfield  {title} {\bibinfo {title} {Tamper-{Indicating} {Quantum} {Seal}},\ }\href {https://doi.org/10.1103/physrevapplied.5.014001} {\bibfield  {journal} {\bibinfo  {journal} {Phys. Rev. Applied}\ }\textbf {\bibinfo {volume} {5}},\ \bibinfo {pages} {014001} (\bibinfo {year} {2016})}\BibitemShut {NoStop}%
\bibitem [{\citenamefont {Zhao}\ \emph {et~al.}(2022)\citenamefont {Zhao}, \citenamefont {Lyons}, \citenamefont {Ulku}, \citenamefont {Defienne}, \citenamefont {Faccio},\ and\ \citenamefont {Charbon}}]{ZhaoOE2022}%
  \BibitemOpen
  \bibfield  {author} {\bibinfo {author} {\bibfnamefont {J.}~\bibnamefont {Zhao}}, \bibinfo {author} {\bibfnamefont {A.}~\bibnamefont {Lyons}}, \bibinfo {author} {\bibfnamefont {A.~C.}\ \bibnamefont {Ulku}}, \bibinfo {author} {\bibfnamefont {H.}~\bibnamefont {Defienne}}, \bibinfo {author} {\bibfnamefont {D.}~\bibnamefont {Faccio}},\ and\ \bibinfo {author} {\bibfnamefont {E.}~\bibnamefont {Charbon}},\ }\bibfield  {title} {\bibinfo {title} {Light detection and ranging with entangled photons},\ }\href {https://doi.org/10.1364/oe.435898} {\bibfield  {journal} {\bibinfo  {journal} {Opt. Express}\ }\textbf {\bibinfo {volume} {30}},\ \bibinfo {pages} {3675} (\bibinfo {year} {2022})}\BibitemShut {NoStop}%
\bibitem [{\citenamefont {Oh}\ \emph {et~al.}(2023)\citenamefont {Oh}, \citenamefont {Jiang},\ and\ \citenamefont {Fefferman}}]{OhPRL2023}%
  \BibitemOpen
  \bibfield  {author} {\bibinfo {author} {\bibfnamefont {C.}~\bibnamefont {Oh}}, \bibinfo {author} {\bibfnamefont {L.}~\bibnamefont {Jiang}},\ and\ \bibinfo {author} {\bibfnamefont {B.}~\bibnamefont {Fefferman}},\ }\bibfield  {title} {\bibinfo {title} {Spoofing cross-entropy measure in boson sampling},\ }\href {https://doi.org/10.1103/physrevlett.131.010401} {\bibfield  {journal} {\bibinfo  {journal} {Phys. Rev. Lett.}\ }\textbf {\bibinfo {volume} {131}},\ \bibinfo {pages} {010401} (\bibinfo {year} {2023})}\BibitemShut {NoStop}%
\bibitem [{\citenamefont {Blakely}\ and\ \citenamefont {Pethel}(2022)}]{BlakelyPRR2022}%
  \BibitemOpen
  \bibfield  {author} {\bibinfo {author} {\bibfnamefont {J.~N.}\ \bibnamefont {Blakely}}\ and\ \bibinfo {author} {\bibfnamefont {S.~D.}\ \bibnamefont {Pethel}},\ }\bibfield  {title} {\bibinfo {title} {Quantum limits to classically spoofing an electromagnetic signal},\ }\href {https://doi.org/10.1103/physrevresearch.4.023178} {\bibfield  {journal} {\bibinfo  {journal} {Phys. Rev. Res.}\ }\textbf {\bibinfo {volume} {4}},\ \bibinfo {pages} {023178} (\bibinfo {year} {2022})}\BibitemShut {NoStop}%
\bibitem [{\citenamefont {Schleher}(1999)}]{Schleher}%
  \BibitemOpen
  \bibfield  {author} {\bibinfo {author} {\bibfnamefont {D.~C.}\ \bibnamefont {Schleher}},\ }\href@noop {} {\emph {\bibinfo {title} {Electronic Warfare in the Information Age}}},\ \bibinfo {edition} {1st}\ ed.\ (\bibinfo  {publisher} {Artech House, Inc.},\ \bibinfo {address} {USA},\ \bibinfo {year} {1999})\BibitemShut {NoStop}%
\bibitem [{\citenamefont {Blakely}\ \emph {et~al.}(2024)\citenamefont {Blakely}, \citenamefont {Pethel},\ and\ \citenamefont {Jacobs}}]{BlakelyPRR2024}%
  \BibitemOpen
  \bibfield  {author} {\bibinfo {author} {\bibfnamefont {J.~N.}\ \bibnamefont {Blakely}}, \bibinfo {author} {\bibfnamefont {S.~D.}\ \bibnamefont {Pethel}},\ and\ \bibinfo {author} {\bibfnamefont {K.}~\bibnamefont {Jacobs}},\ }\bibfield  {title} {\bibinfo {title} {Revealing spoofing of classical radar using quantum noise},\ }\href {https://doi.org/10.1103/physrevresearch.6.013179} {\bibfield  {journal} {\bibinfo  {journal} {Phys. Rev. Res.}\ }\textbf {\bibinfo {volume} {6}},\ \bibinfo {pages} {013179} (\bibinfo {year} {2024})}\BibitemShut {NoStop}%
\bibitem [{\citenamefont {Senellart}\ \emph {et~al.}(2017)\citenamefont {Senellart}, \citenamefont {Solomon},\ and\ \citenamefont {White}}]{SenellartNN2017}%
  \BibitemOpen
  \bibfield  {author} {\bibinfo {author} {\bibfnamefont {P.}~\bibnamefont {Senellart}}, \bibinfo {author} {\bibfnamefont {G.}~\bibnamefont {Solomon}},\ and\ \bibinfo {author} {\bibfnamefont {A.}~\bibnamefont {White}},\ }\bibfield  {title} {\bibinfo {title} {High-performance semiconductor quantum-dot single-photon sources},\ }\href {https://doi.org/10.1038/nnano.2017.218} {\bibfield  {journal} {\bibinfo  {journal} {Nature Nanotech}\ }\textbf {\bibinfo {volume} {12}},\ \bibinfo {pages} {1026} (\bibinfo {year} {2017})}\BibitemShut {NoStop}%
\bibitem [{\citenamefont {Cook}\ \emph {et~al.}(2007)\citenamefont {Cook}, \citenamefont {Martin},\ and\ \citenamefont {Geremia}}]{CookN2007}%
  \BibitemOpen
  \bibfield  {author} {\bibinfo {author} {\bibfnamefont {R.~L.}\ \bibnamefont {Cook}}, \bibinfo {author} {\bibfnamefont {P.~J.}\ \bibnamefont {Martin}},\ and\ \bibinfo {author} {\bibfnamefont {J.~M.}\ \bibnamefont {Geremia}},\ }\bibfield  {title} {\bibinfo {title} {Optical coherent state discrimination using a closed-loop quantum measurement},\ }\href {https://doi.org/10.1038/nature05655} {\bibfield  {journal} {\bibinfo  {journal} {Nature}\ }\textbf {\bibinfo {volume} {446}},\ \bibinfo {pages} {774} (\bibinfo {year} {2007})}\BibitemShut {NoStop}%
\bibitem [{\citenamefont {Wittmann}\ \emph {et~al.}(2008)\citenamefont {Wittmann}, \citenamefont {Takeoka}, \citenamefont {Cassemiro}, \citenamefont {Sasaki}, \citenamefont {Leuchs},\ and\ \citenamefont {Andersen}}]{WittmannPRL2008}%
  \BibitemOpen
  \bibfield  {author} {\bibinfo {author} {\bibfnamefont {C.}~\bibnamefont {Wittmann}}, \bibinfo {author} {\bibfnamefont {M.}~\bibnamefont {Takeoka}}, \bibinfo {author} {\bibfnamefont {K.~N.}\ \bibnamefont {Cassemiro}}, \bibinfo {author} {\bibfnamefont {M.}~\bibnamefont {Sasaki}}, \bibinfo {author} {\bibfnamefont {G.}~\bibnamefont {Leuchs}},\ and\ \bibinfo {author} {\bibfnamefont {U.~L.}\ \bibnamefont {Andersen}},\ }\bibfield  {title} {\bibinfo {title} {Demonstration of near-optimal discrimination of optical coherent states},\ }\href {https://doi.org/10.1103/physrevlett.101.210501} {\bibfield  {journal} {\bibinfo  {journal} {Phys. Rev. Lett.}\ }\textbf {\bibinfo {volume} {101}},\ \bibinfo {pages} {210501} (\bibinfo {year} {2008})}\BibitemShut {NoStop}%
\bibitem [{\citenamefont {Ferdinand}\ \emph {et~al.}(2017)\citenamefont {Ferdinand}, \citenamefont {DiMario},\ and\ \citenamefont {Becerra}}]{FerdinandNPJQI2017}%
  \BibitemOpen
  \bibfield  {author} {\bibinfo {author} {\bibfnamefont {A.~R.}\ \bibnamefont {Ferdinand}}, \bibinfo {author} {\bibfnamefont {M.~T.}\ \bibnamefont {DiMario}},\ and\ \bibinfo {author} {\bibfnamefont {F.~E.}\ \bibnamefont {Becerra}},\ }\bibfield  {title} {\bibinfo {title} {Multi-state discrimination below the quantum noise limit at the single-photon level},\ }\href {https://doi.org/10.1038/s41534-017-0042-2} {\bibfield  {journal} {\bibinfo  {journal} {npj Quantum Inf}\ }\textbf {\bibinfo {volume} {3}},\ \bibinfo {pages} {43} (\bibinfo {year} {2017})}\BibitemShut {NoStop}%
\bibitem [{\citenamefont {Schnabel}(2017)}]{SchnabelPR2017}%
  \BibitemOpen
  \bibfield  {author} {\bibinfo {author} {\bibfnamefont {R.}~\bibnamefont {Schnabel}},\ }\bibfield  {title} {\bibinfo {title} {Squeezed states of light and their applications in laser interferometers},\ }\href {https://doi.org/10.1016/j.physrep.2017.04.001} {\bibfield  {journal} {\bibinfo  {journal} {Phys. Rep.}\ }\textbf {\bibinfo {volume} {684}},\ \bibinfo {pages} {1} (\bibinfo {year} {2017})}\BibitemShut {NoStop}%
\bibitem [{\citenamefont {Walls}(1983)}]{WallsN1983}%
  \BibitemOpen
  \bibfield  {author} {\bibinfo {author} {\bibfnamefont {D.~F.}\ \bibnamefont {Walls}},\ }\bibfield  {title} {\bibinfo {title} {Squeezed states of light},\ }\href {https://doi.org/10.1038/306141a0} {\bibfield  {journal} {\bibinfo  {journal} {Nature}\ }\textbf {\bibinfo {volume} {306}},\ \bibinfo {pages} {141} (\bibinfo {year} {1983})}\BibitemShut {NoStop}%
\bibitem [{\citenamefont {Slusher}\ \emph {et~al.}(1985)\citenamefont {Slusher}, \citenamefont {Hollberg}, \citenamefont {Yurke}, \citenamefont {Mertz},\ and\ \citenamefont {Valley}}]{SlusherPRL1985}%
  \BibitemOpen
  \bibfield  {author} {\bibinfo {author} {\bibfnamefont {R.~E.}\ \bibnamefont {Slusher}}, \bibinfo {author} {\bibfnamefont {L.~W.}\ \bibnamefont {Hollberg}}, \bibinfo {author} {\bibfnamefont {B.}~\bibnamefont {Yurke}}, \bibinfo {author} {\bibfnamefont {J.~C.}\ \bibnamefont {Mertz}},\ and\ \bibinfo {author} {\bibfnamefont {J.~F.}\ \bibnamefont {Valley}},\ }\bibfield  {title} {\bibinfo {title} {Observation of squeezed states generated by four-wave mixing in an optical cavity},\ }\href {https://doi.org/10.1103/physrevlett.55.2409} {\bibfield  {journal} {\bibinfo  {journal} {Phys. Rev. Lett.}\ }\textbf {\bibinfo {volume} {55}},\ \bibinfo {pages} {2409} (\bibinfo {year} {1985})}\BibitemShut {NoStop}%
\bibitem [{\citenamefont {Ou}(2017)}]{ou2017quantum}%
  \BibitemOpen
  \bibfield  {author} {\bibinfo {author} {\bibfnamefont {Z.~J.}\ \bibnamefont {Ou}},\ }\href@noop {} {\emph {\bibinfo {title} {Quantum optics for experimentalists}}}\ (\bibinfo  {publisher} {World Scientific Publishing Company},\ \bibinfo {year} {2017})\BibitemShut {NoStop}%
\bibitem [{\citenamefont {Vahlbruch}\ \emph {et~al.}(2016)\citenamefont {Vahlbruch}, \citenamefont {Mehmet}, \citenamefont {Danzmann},\ and\ \citenamefont {Schnabel}}]{VahlbruchPRL2016}%
  \BibitemOpen
  \bibfield  {author} {\bibinfo {author} {\bibfnamefont {H.}~\bibnamefont {Vahlbruch}}, \bibinfo {author} {\bibfnamefont {M.}~\bibnamefont {Mehmet}}, \bibinfo {author} {\bibfnamefont {K.}~\bibnamefont {Danzmann}},\ and\ \bibinfo {author} {\bibfnamefont {R.}~\bibnamefont {Schnabel}},\ }\bibfield  {title} {\bibinfo {title} {Detection of 15 db squeezed states of light and their application for the absolute calibration of photoelectric quantum efficiency},\ }\href {https://doi.org/10.1103/physrevlett.117.110801} {\bibfield  {journal} {\bibinfo  {journal} {Phys. Rev. Lett.}\ }\textbf {\bibinfo {volume} {117}},\ \bibinfo {pages} {110801} (\bibinfo {year} {2016})}\BibitemShut {NoStop}%
\bibitem [{\citenamefont {Møller}\ \emph {et~al.}(1996)\citenamefont {Møller}, \citenamefont {Jørgensen},\ and\ \citenamefont {Dahl}}]{mo1996displaced}%
  \BibitemOpen
  \bibfield  {author} {\bibinfo {author} {\bibfnamefont {K.~B.}\ \bibnamefont {Møller}}, \bibinfo {author} {\bibfnamefont {T.~G.}\ \bibnamefont {Jørgensen}},\ and\ \bibinfo {author} {\bibfnamefont {J.~P.}\ \bibnamefont {Dahl}},\ }\bibfield  {title} {\bibinfo {title} {Displaced squeezed number states: Position space representation, inner product, and some applications},\ }\href {https://doi.org/10.1103/physreva.54.5378} {\bibfield  {journal} {\bibinfo  {journal} {Phys. Rev. A}\ }\textbf {\bibinfo {volume} {54}},\ \bibinfo {pages} {5378} (\bibinfo {year} {1996})}\BibitemShut {NoStop}%
\end{thebibliography}%

\end{document}